\documentclass[preprint]{aastex6}

\begin{document}                                                                     


\title{X-ray Observations of the Peculiar Cepheid V473 Lyr Identify 
A Low--Mass Companion \altaffilmark{1},
 }


\author{
Nancy Remage Evans,\altaffilmark{2}
Ignazio Pillitteri,\altaffilmark{3} 
Laszlo Molnar,\altaffilmark{4}
Laszlo Szabados,\altaffilmark{4}
Emese Plachy,\altaffilmark{4}
Robert Szabo,\altaffilmark{4}
Scott  Engle,\altaffilmark{5} 
Edward Guinan,\altaffilmark{5} 
Scott Wolk,\altaffilmark{2} 
H. Moritz G\"unther,\altaffilmark{6}
Hilding Neilson,\altaffilmark{7}
Massimo Marengo,\altaffilmark{8}
Lynn D. Matthews,\altaffilmark{9}
Sofia Moschou,\altaffilmark{2}
Jeremy J. Drake,\altaffilmark{2}
Vinay Kashyap,\altaffilmark{2}
Pierre Kervella,\altaffilmark{10} 
 Tamas Tordai,\altaffilmark{11}
Peter Somogyi,\altaffilmark{12}
and Gilbert Burki\altaffilmark{13}
}  


\altaffiltext{1}
{Based on observations obtained with  {\it XMM-Newton}, an 
ESA science mission with instruments and contributions directly 
funded by ESA Member States and the USA (NASA)}




\altaffiltext{2}
{Smithsonian Astrophysical Observatory,    
MS 4, 60 Garden St., Cambridge, MA 02138; nevans@cfa.harvard.edu}

\altaffiltext{3}
{INAF-Osservatorio di Palermo, Piazza del Parlamento 1, 90134 Palermo, Italy}

\altaffiltext{4}
{Konkoly Observatory, and 
MTA CSFK Lend\"ulet Near-Field Cosmology Research Group, 
 Konkoly Thege Miklos ut 15-17, H-1121
Budapest, Hungary}






\altaffiltext{5}
{Department of Astronomy and Astrophysics, Villanova University, 800 Lancaster Ave., 
Villanova, PA, 19085, USA}

\altaffiltext{6}
{Massachusetts Institute of Technology, Kavli Institute for Astrophysics and
Space Research, 77 Massachusetts Ave, NE83-569, Cambridge MA 02139, USA}

\altaffiltext{7}
{Department of Astronomy and Astrophysics, University of Toronto, 
50 St. George Street, Toronto, ON, Canada M5S3H4}

\altaffiltext{8}
{Department of Physics and Astronomy, Iowa State University, Ames, IA, 50011,
 USA}

\altaffiltext{9}
{Massachusetts Institute of Technology, Haystack Observatory, 99 Millstone Rd., 
Westford, MA 01886, USA}

\altaffiltext{10}
{LESIA, Observatoire de Paris, Universit\'e PSL, CNRS, Sorbonne Universit\'e, Univ. Paris Diderot, Sorbonne Paris Cit\'e, 5 place Jules Janssen, 92195 Meudon, France}

\altaffiltext{11}
{ Polaris Observatory, Hungarian Astronomical Association, Laborc utca 
2/c, 1037 Budapest, Hungary}

\altaffiltext{12}
{Zrinyi u. 23., 2890, Tata, Hungary}

\altaffiltext{13}
{Observatoire de Genève, 51 chemin des Maillettes, 
1290 Sauverny, Switzerland}





\begin{abstract}
V473 Lyr is a classical Cepheid which is unique in having 
substantial amplitude variations with a period of approximately
3.3 years, thought to be similar to the Blazhko variations in 
RR Lyrae stars.  We obtained an {\it XMM-Newton} observation of this
star to followup a previous detection in X-rays.  Rather than 
the X-ray burst and rapid decline 
near maximum radius seen in $\delta$ Cephei 
itself, the X-ray flux in V473 Lyr remained constant for  a 
third of the pulsation cycle covered by the observation.    
Thus the X-rays are most 
probably not  produced by the changes around the pulsation cycle.  
The X-ray spectrum is soft (kT = 0.6 keV), with 
 X-ray properties which
are consistent with a young low mass companion.  Previously
there was no evidence of a companion in radial velocities or 
in {\it Gaia} and {\it Hipparcos} proper motions. While this
rules out companions which are very close or very distant,
a binary companion at a separation between 30 and 300 AU
 is possible.  This is an example of an X-ray observation 
revealing evidence of 
a low mass companion, which is important in completing
the mass ratio 
statistics of binary Cepheids.  Furthermore, the detection of a 
young X-ray bright
companion is a further indication that the Cepheid (primary)
is a Population I star, even though its pulsation behavior 
differs from other classical Cepheids.

\end{abstract}




\section{Introduction}


Pop I classical Cepheids are radial pulsators 
that have very regular pulsation cycles. Complicated 
photometric variations, however, are found in 
 a small group which are  excited in two modes. Long series 
of precise photometry from satellites such as {\it Kepler}, 
{\it MOST}, and {\it CoRoT} have begun to alert us to 
additional excited frequencies, particularly in overtone 
pulsators.   The notable 
exception to the regular pulsation is V473 Lyr, which 
has a variable amplitude of pulsation.  The main pulsation 
period is 1.49$^d$, with a period of amplitude variation of 
1205$^d$.  Existing data are discussed by Molnar and Szabados
(2014; MS below).  This amplitude variation appears to 
be similar to the Blazhko effect seen in many RR Lyr stars.
As discussed there, V473 Lyr is thought to be 
pulsating in the second overtone.  A number of explanations 
have been put forward for the amplitude variation, but the 
most likely is the resonance between two pulsation modes, which is discussed
by MS.  They identified an additional modulation cycle of 
5300$^d$. A sequence of 27 days of observations with the 
{\it MOST} satellite (Molnar, et al. 2017)
shows period doubling (the alternation of the amplitudes of
successive cycles).  This is seen in RV Tau and RR Lyr Pop II 
pulsators, but this is the first case in a Pop I classical 
Cepheid.

X-ray observations of classical Cepheids have a peculiar
pattern (Engle, et al. 2017).  The pulsation cycle results in 
disturbances in the photosphere and chromosphere following
minimum radius as the pulsation wave passes through.  
X-ray flux of $\delta$ Cep is relatively modest 
and constant at this phase.  Just after maximum radius, 
however, the X-ray flux increases sharply by a factor of 
approximately four. At this phase,  photospheric and   
chromospheric spectra are quiescent, indistinguishable 
from those of nonvariable supergiants.  
This pattern is seen in two cycles of 
$\delta$ Cep and  in $\beta$ Dor.


V473 Lyr was observed by the {\it XMM} satellite in a program 
to observe 14 Cepheids (Evans, et al. 2016a),  
which  was developed from  observations 
using the {\it Hubble Space Telescope (HST)} Widefield Camera 3
(WFC3)  to identify possible resolved companions  
 (Evans, et al. 2016b).   The {\it XMM} observations 
were used to distinguish X-ray active low mass stars young 
enough to be Cepheid companions from old field stars which 
are much less active.  V473 Lyr was found to have an X-ray source at the 
position of the Cepheid.   
Table 1 lists the phase of the observation based 
on complex pulsation behavior discussed in MS.
 The phase of the 
X-ray observation was found to be the same as 
the phase at which $\delta$ Cep has a burst of X-rays, namely
just after maximum radius (Evans, et al. 2018).  This suggests
the possibility 
 that the X-ray flux was produced in the same way
as for $\delta$ Cep.  For this reason 
an additional {\it XMM} observation was 
applied for, partly to use the variation in pulsation 
amplitude as a diagnostic of the effect.  

\begin{deluxetable}{lllrr}
\tablecaption{{\it XMM} Observations of V473 Lyr  \qquad\qquad\qquad\qquad \label{}}
\tablewidth{0pt}
\tablehead{
\colhead{Year} &  \colhead{}  & \colhead{JD}  & \colhead{Exp}  & \colhead{Phase} \\
 \colhead{}  &  \colhead{} &  \colhead{$-$2400000} &  \colhead{ksec} &  \colhead{} \\
}
\startdata
2019  & Start & 58559.762 & 40.3  &  0.42  \\
      &  End & 58560.228  &   & 0.73   \\
 & & & & \\
2013  & Start & 56557.909 & 6.6 & 0.47 \\
      & End  & 56558.002  &   & 0.53 \\
\enddata

\end{deluxetable}

\section{Observation and Data Analysis}

\subsection{{\it XMM} Observation}

A new observation of V473 Lyr with {\it XMM} was obtained 
in 2019 
(Table 1).  
Data analysis was carried out using standard data reduction 
tasks in SAS software (Scientific Analysis Subsystem) version 17.0
 as in Pillitteri, et al. (2013).  This involved a reduction 
starting from the ODFs of the observation, filtering the events 
according to their grades
and screening out bad pixels. Only events between 0.3 and 8.0 keV 
were used, and
the reduction was restricted to good time intervals and low 
background periods.
These  were evaluated using the recipe given in the SAS 
guidelines and based on the light curve of the events above 10 keV.

\begin{figure}
\plotone{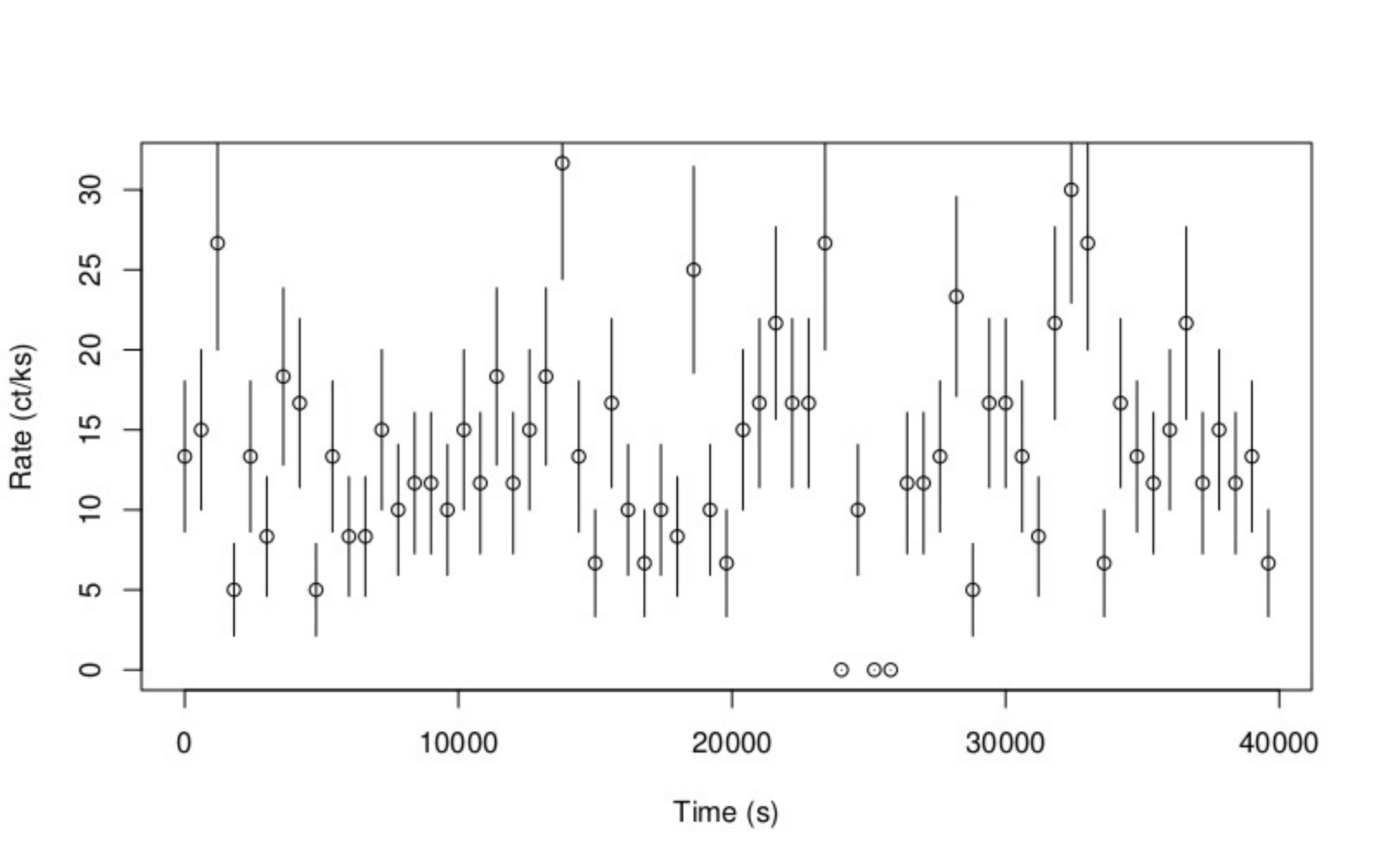}
\caption{ Count rate during the {\it XMM} exposure.  
\label{cts}}
\end{figure}
\begin{figure}
\plotone{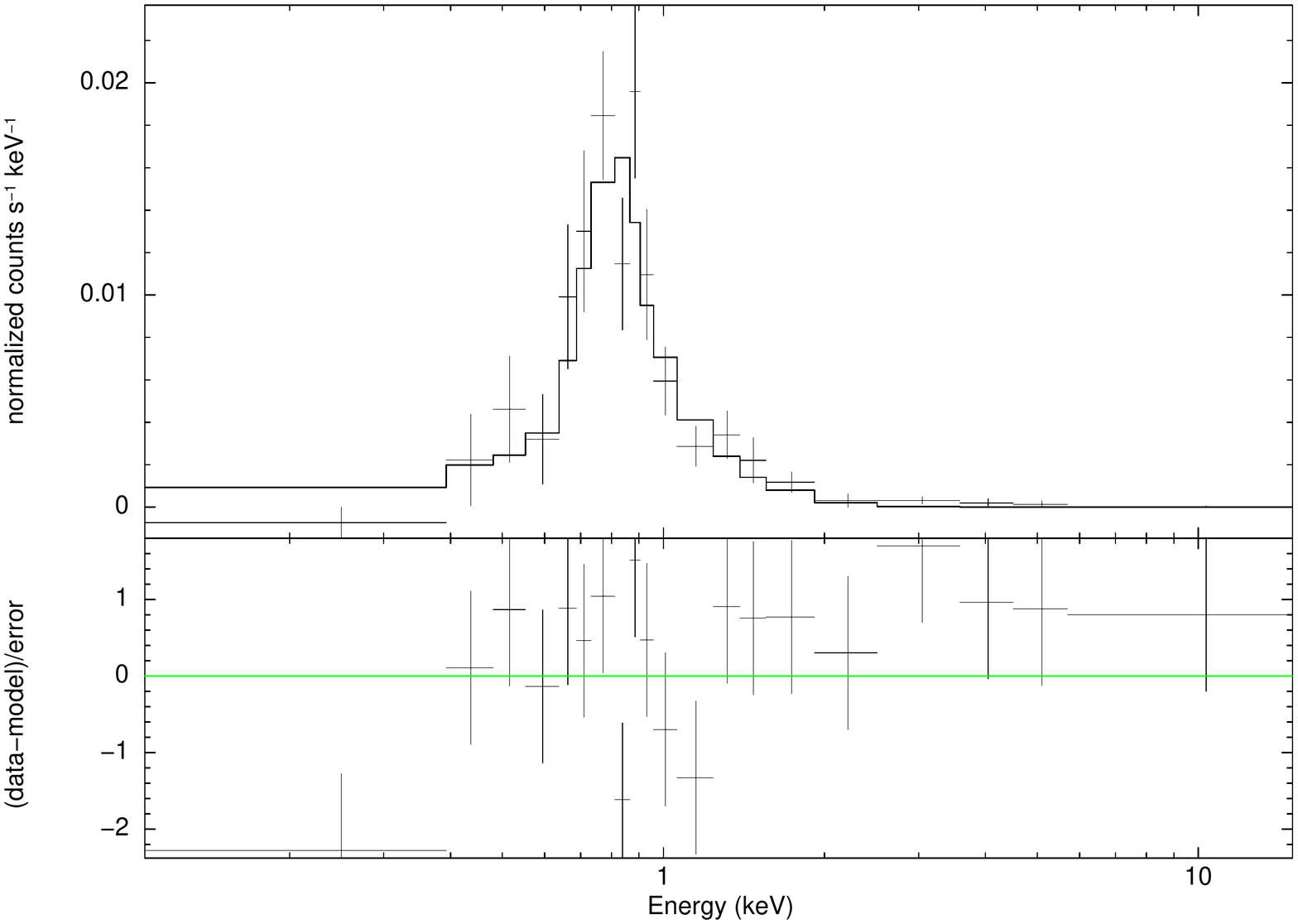}
\caption{ Extracted 
spectrum of V473 Lyr.  Top: +'s: data; histogram: data fit.  Bottom: residuals.
\label{spect}}
\end{figure}

Again there was a source at the position of the Cepheid.  The 
light variation (PN chip array) and the spectrum are shown in 
Fig.~\ref{cts} and Fig.~\ref{spect}.  
Fig.~\ref{cts} shows that the 
flux is essentially constant during the exposure.  Fig~\ref{spect}
 shows the spectrum  fit given for a 
temperature of kT = 0.6 keV (using a H column density 
of 3 $\times$  10$^{21}$ cm$^{-2}$ [appropriate for a low E(B-V) = 0.03 mag] 
and an abundance of 0.3 Z$_\odot$).  
The unabsorbed flux is 3.2  $\times$ 10$^{-14}$ erg cm$^{-2}$ s$^{-1}$ 
 in the 0.3 to 8.0 keV band.  
Using a distance of 553 pc (Evans, et al. 2016b), 
this is log Lx = 30.07  erg s$^{-1}$.  This distance is 
based on the Benedict, et al. (2007) Leavitt Law.  
This is  larger than the distance 
calculated from the data in the {\it Gaia} DR2  
catalog ({\it Gaia} Collaboration et al. 2018). 

    We have tested whether variability is detectable in the light curve
Fig.~\ref{cts}
using the observed fluctuations in the counts against a constant count
    rate model.  We account for the presence of a constant background rate,
    and exclude times when the PN background flares.  We compute
    a {\tt cstat} value (Cash statistic; Cash 1979)
and corresponding approximate goodness-of-fit
    measure (Kaastra 2017) for the data compared with the averaged
    counts in light curves binned  from 100 to 1000 s.
We have also compared the histogram of counts from a binned 
counts light curve 
against the predicted Poisson frequency. 
The PN data have the most counts, but we have also examined 
MOS1 and MOS2 both separately and together. 
    The results varied depending on the  detector and time bin,  but
we find marginal evidence for variability at the largest 
binning at the 10\% level.

\subsection{Photometry}

\begin{figure}
\plotone{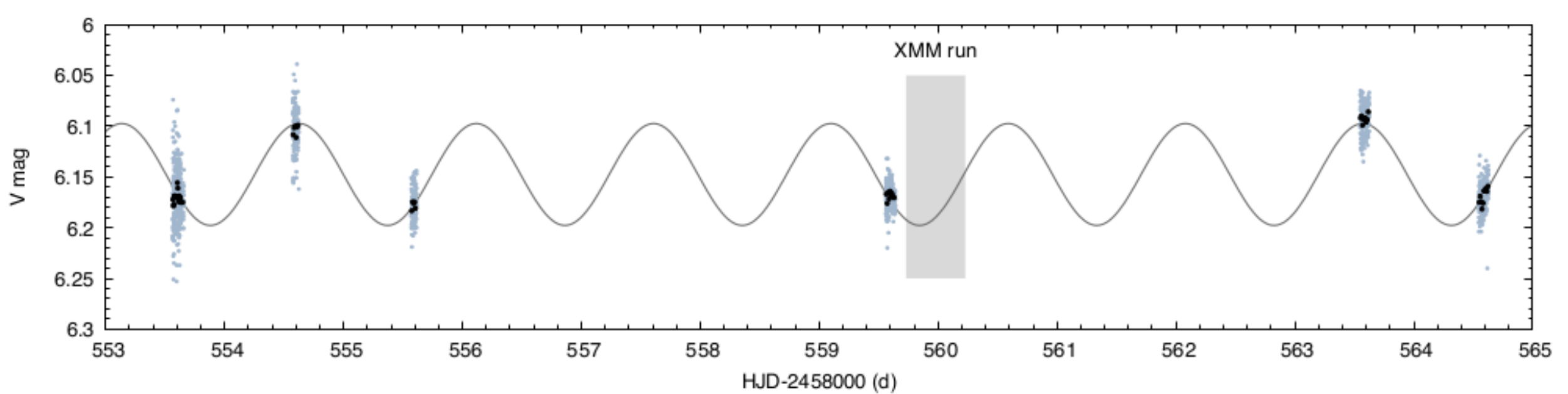}
\caption{New photometric data and 
the phases of the {\it XMM} observation. 
Blue dots: V magnitude data; black dots: 20 point binned 
data; solid line: pulsation curve 
computed from the ephemeris; shaded region: the 
time {\it XMM} of the observation.
\label{obs.phas}
}
\end{figure}

Since the pulsation amplitude of V473 Lyr is variable,  photometry was 
obtained near the time of the {\it XMM}  
observation to confirm the phases.
Observations were made  near Budapest, Hungary,
 with a 15 cm telescope, and an Orion StarShoot G3 CCD camera, in the V band. 
During the first night, defocused images were taken with 5 sec integrations 
to avoid saturation. For the second night the telescope was used with an 
aperture opening reduced to 4 cm diameter, with 30 sec integration times. 
HD 180316 (V = 6.891 mag) and HD 337922 (V = 9.171 mag) were used as 
comparison and check stars. Differential magnitudes were shifted to the 
average brightness of the star (V = 6.153 mag), and times were converted 
to HJD.  Data are listed in Appendix A (full table in the electronic
version).
The data were binned  (20 points) to decrease the scatter. Since
 the pulsation at this time
is at a low-amplitude phase and the light curve is nearly sinusoidal, 
we  fitted the observations with a simple sine, with the frequency 
fixed to the average pulsation frequency value (0.67079 d$^{-1}$). This way 
we were able to confirm the pulsation phase during the observing window
 for XMM-Newton. As Fig.~\ref{obs.phas} shows, data collection started 
slightly before minimum light, and ended about halfway up the ascending 
branch.





\begin{figure}
\plotone{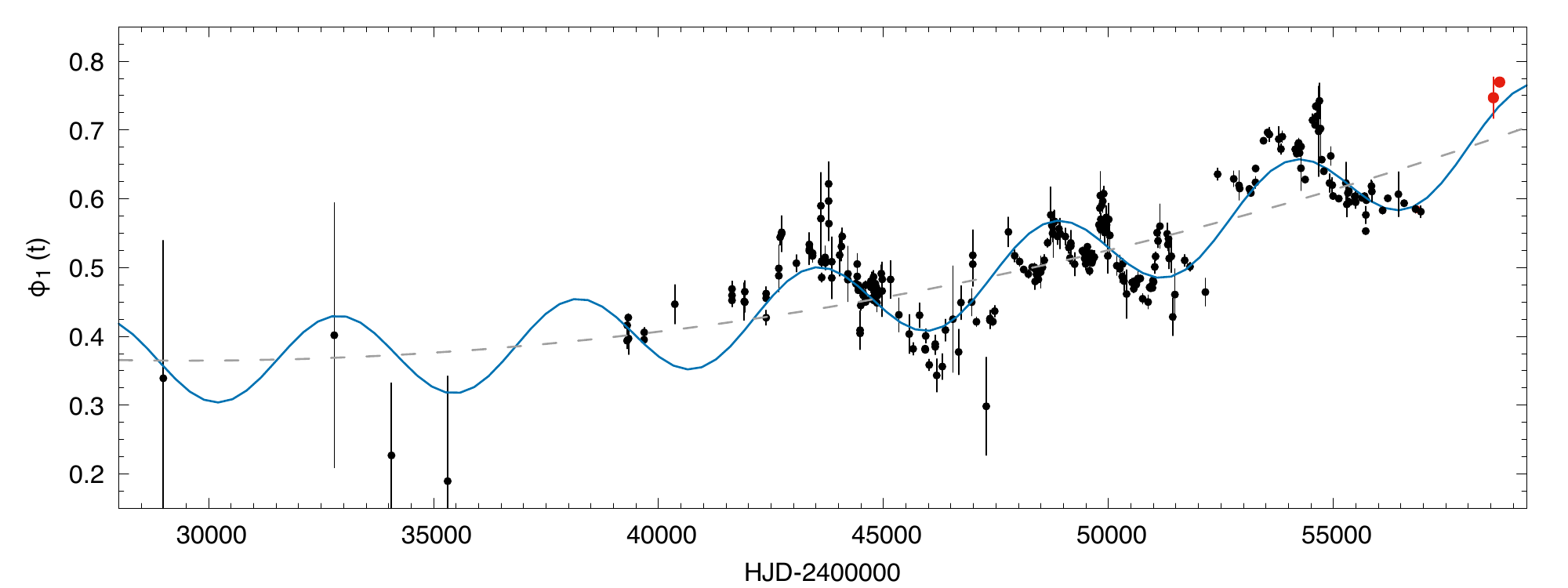}
\caption{Long-term phase variation V473 Lyrae, based on the $\phi_1$ Fourier term. The dashed line is a quadratic term describing a continuous decrease in period. The blue line shows the longer modulation cycle superimposed. This plot is an updated version of Fig. 9 by Molnar \& Szabados (2014), and includes photometric data from Molnar et al. (2017), and Appendix A. The red dots on the far 
right are  from Appendix A and the TESS data. 
\label{obs.long.phas}
}
\end{figure}

Fig.~\ref{obs.long.phas} shows the updated phase 
variation of V473 Lyr, with the photometry from Fig.~\ref{obs.phas}
included.  

\subsection{Velocities: 2019}

Radial velocity data were collected on two nights from Hungary with 
a 30 cm Newton telescope, and a Shelyak LHires III spectrograph, 
at R 13000 resolution, on 17th March, 2019, right before the 
XMM run started, and again on the 24th, for 2.3 and 2.7 hours, 
respectively. An Ar/Ne calibration lamp and the radial velocity standard 
star 45 Dra were used for wavelength calibration. Images were 
processed with ISIS\footnote {http://www.astrosurf.com/buil/isis-software.html},
 then the spectra were rectified with 
IRAF\footnote {http://ui.adsabs.harvard.edu/abs/1986SPIE..627..733T}. 
Lines between 6103 and 6223 \AA{} were used to determine the 
radial velocities  with the IRAF cross-correlation task 
fxcor. Uncertainties for the target and standard star measurements 
were both estimated to be 0.5 km/s. Velocities are listed in
Appendix B (full table in the electronic version).

\subsection{{\it TESS} Observations}

\begin{figure}
\plotone{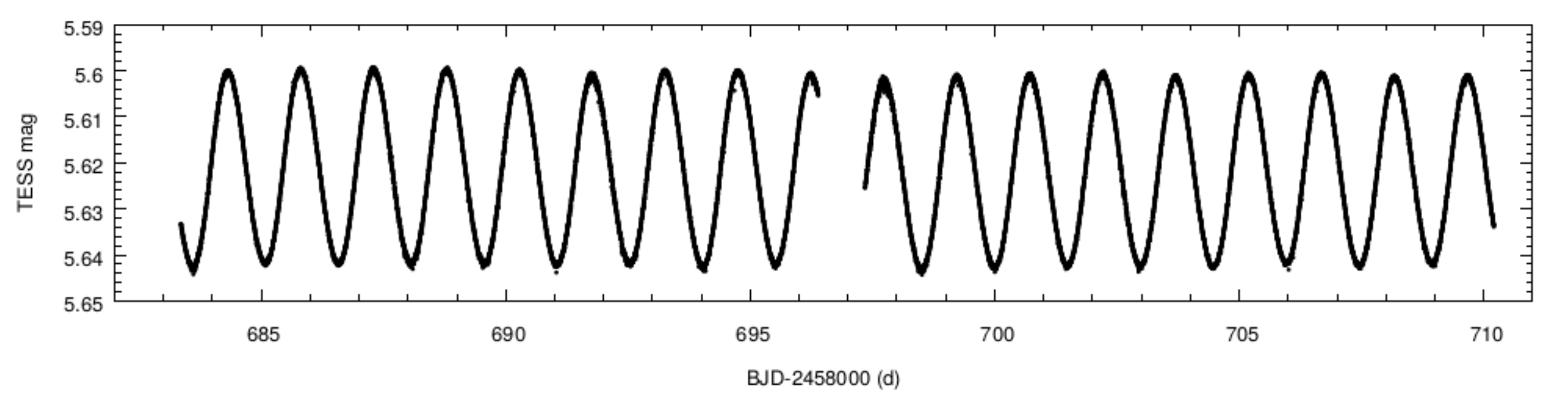}
\caption{{\it TESS} observations of V473 Lyr. 
\label{obs.tess}
}
\end{figure}

V473 Lyr was observed by the TESS space telescope in Sector 14, a few months 
after the XMM run, from July 18th to August 14, 2019 (Ricker et al., 2014). 
Observations were made in short cadence mode, with 2 min sampling. We 
analyzed the light curve released by the Spacecraft Operations Center 
(SPOC) and created custom-aperture photometry with our \emph{lightkurve} 
package (Lightkurve Collaboration, 2018 
\footnote {http://adsabs.harvard.edu/abs/2018ascl.soft12013L})
which is shown in Fig. ~\ref{obs.tess}.
Since the star 
is in a low-amplitude phase and TESS observes in a broad red bandpass 
(between 600-1000 nm), the peak-to-peak pulsation amplitude was very 
small, only 41 mmag. We detected the main pulsation peak (f), its first 
harmonic, and one side peak that is connected with the modulation, along 
with some low-frequency components that are likely caused by blending 
with nearby stars and other noise sources.  We 
looked for signs of period doubling, and we found a low-amplitude 
frequency component at 1.5f, but given the presence of low-frequency 
noise in the frequency spectrum nearby, we cannot claim unambiguous detection. 
However, this is not necessarily at odds with the earlier results, 
for two reasons. First, MOST observed the star in shorter wavelengths 
where Cepheid pulsation amplitudes are higher, making detection easier. 
Second, the amplitude of period doubling was found to fluctuate rather 
erratically in RR Lyrae stars: if V473 Lyr behaves similarly, 
non-detection from a short data set can be expected (Szabo et al. 2010). 
We also looked for any signs of flaring activity in the residual light 
curve, but found none with an upper limit of 1 mmag.

\section{Results}

\begin{figure}
\epsscale{0.60}
\plotone{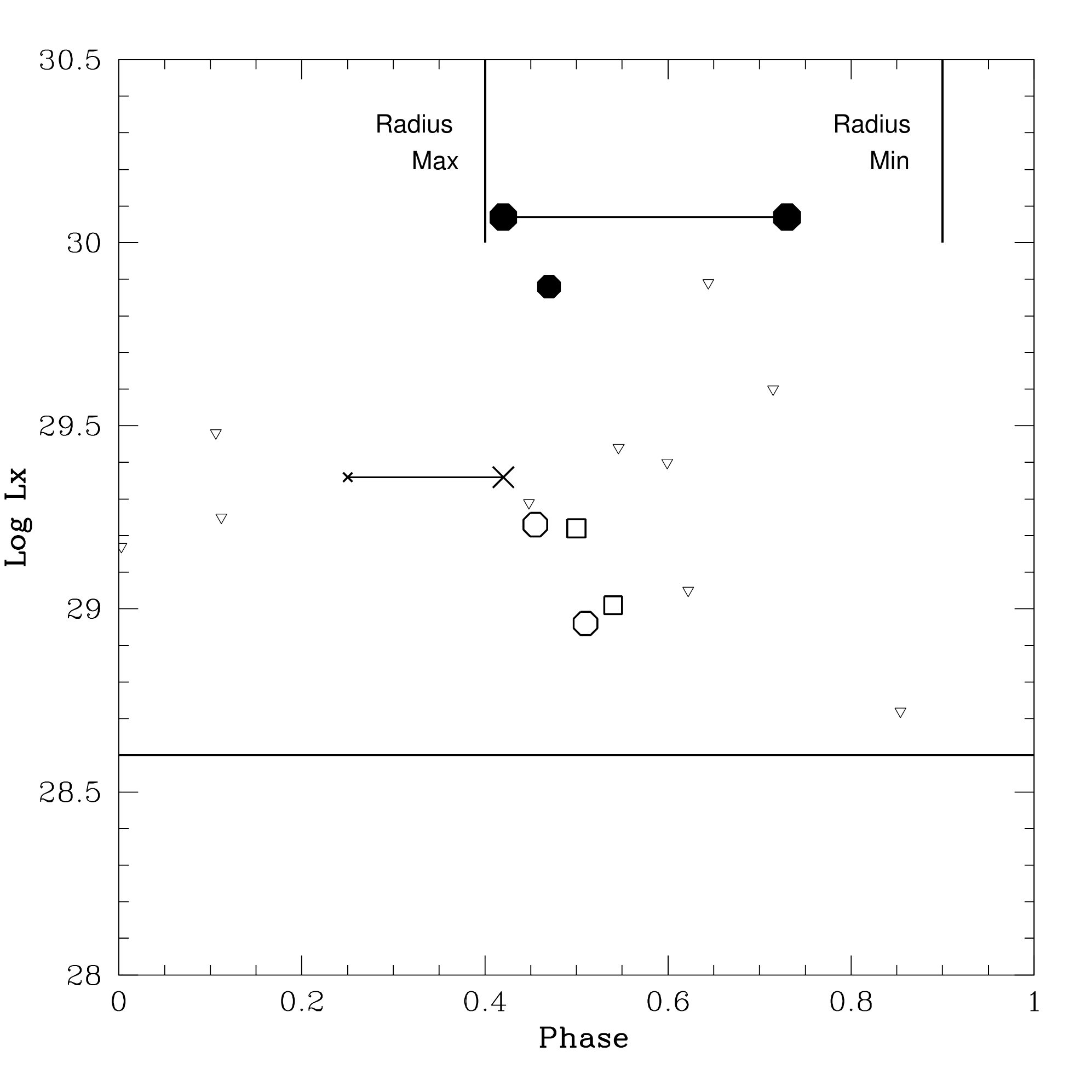}
\caption{Cepheid X-ray observations as a function of pulsation
phase.  Filled circles: observations
of V473 Lyr, with the phase range of the recent observation 
indicated by the joined circles; open circles and squares: $\delta$ Cep,
with different symbols indicating different cycles;
x's: $\beta$ Dor: the phase on the right is adjusted 
to  the phase based on photospheric emission lines (see text);
open triangles:
upper limits for other Cepheids; 
solid line at log L$_X$ = 28.6: $\delta$ Cep in 
quiescent phases. The 2 vertical lines at the top indicate 
maximum and minimum radius of $\delta$ Cep.
 Luminosity is in ergs s$^{-1}$.
\label{cep.sum}}
\end{figure}

The new observation of V473 Lyr is shown in Fig.~\ref{cep.sum}
in the context of other Cepheid X-ray observations.  $\delta$
Cep observations from Engle et al. (2017)  are for both 
quiescent and burst states.  $\beta$ Dor observations include those
from Engle (2015) supplemented by additional observations.   
For $\beta$ Dor, the earlier phase (phase 0.25 on the left in 
Fig.~\ref{cep.sum}) is computed from 
the ephemeris using light maximum.  However, $\beta$ Dor 
is in the period range near 10 days (P = 9.84$^d$) where a
 resonance decreases the pulsation amplitude 
and causes a distortion at maximum light.  Since there are 
{\it HST} COS (Cosmic Origins Spectrograph) 
spectra (Engle 2015) the phase of minimum 
radius is well defined by ultraviolet emission lines caused
by the passage of the pulsation wave through the 
photosphere and chromosphere.
We can use this as the pulsation fiducial
for the relation of minimum radius to the X-ray increase.
Using this fiducial, the phase of the X-ray increase 
for $\beta$ Dor is shifted to the righthand x symbol
in Fig~\ref{cep.sum}.
Upper limits for random phase Cepheid observations are 
from Evans et al. (2016a).   Fig.~\ref{cep.sum} shows that 
the X-ray observations for two  cycles of $\delta$ Cep 
and also $\beta$ Dor in the phase range 0.4 to 0.5 near
maximum radius are well above the quiescent range for other 
phases of $\delta$ Cep.  Both observations of V473 Lyr are 
well above even this bright  X-ray luminosity.  Furthermore
they cover the phase range which extends over a third of the 
pulsation cycle, in contrast to the restricted phase range 
of $\delta$ Cep and $\beta$ Dor.  Although the cause of the 
X-ray bursts in $\delta$ Cep and $\beta$ Dor is not yet 
understood,  the X-ray luminosity
of V473 Lyr does not appear to share the same characteristics
of a much lower quiescent flux for most phases, but a short 
X-ray burst at the phase of maximum radius.

\section{Discussion}

The X-ray source detected for V473 Lyr is relatively constant 
with phase, unlike those seen for $\delta$ Cep and 
$\beta$ Dor, making it much more likely that the X-rays are 
produced by a young low mass companion than pulsation. 
The source is soft (kT = 0.6 keV), which would be
consistent with a G or K main sequence companion at the age of 
the Cepheid (younger than the Pleiades; Preibisch 
and Feigelson 2005).
A cool companion is consistent with the upper limit to a 
companion spectral type of A6 from an {\it IUE (International 
Ultraviolet Explorer)} spectrum
(Evans 1992).  We note that while G or K stars are common in the 
field population, {\it young} G and K stars are very rare, except 
in places such as open clusters.  

\begin{figure}
\plotone{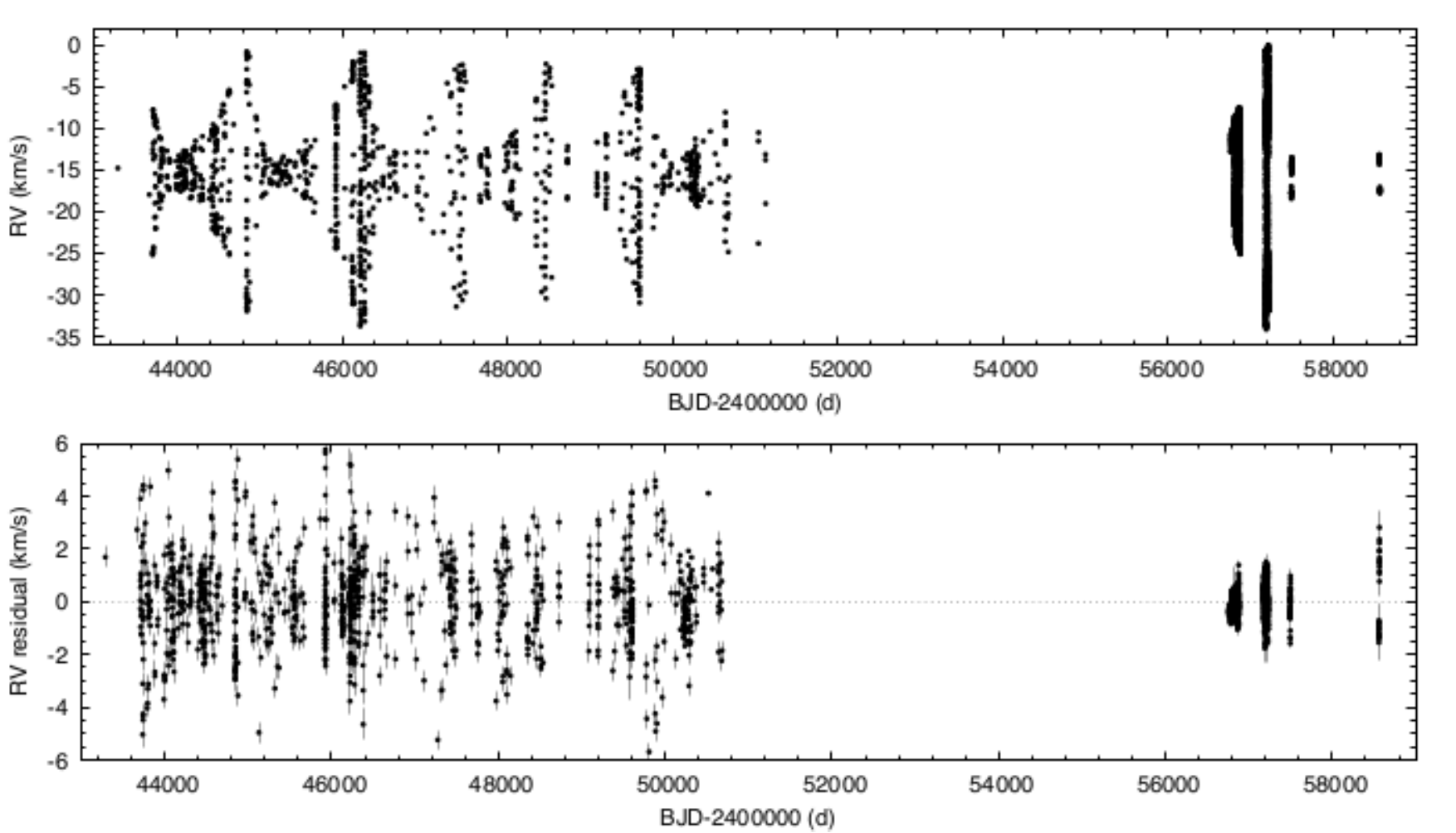}
\caption{Radial velocities of V473 Lyr.  Top: Radial velocities; bottom:
radial velocities corrected for pulsation including modulation
(see text)
\label{vr.burki}}
\end{figure}

\begin{figure}
\plottwo{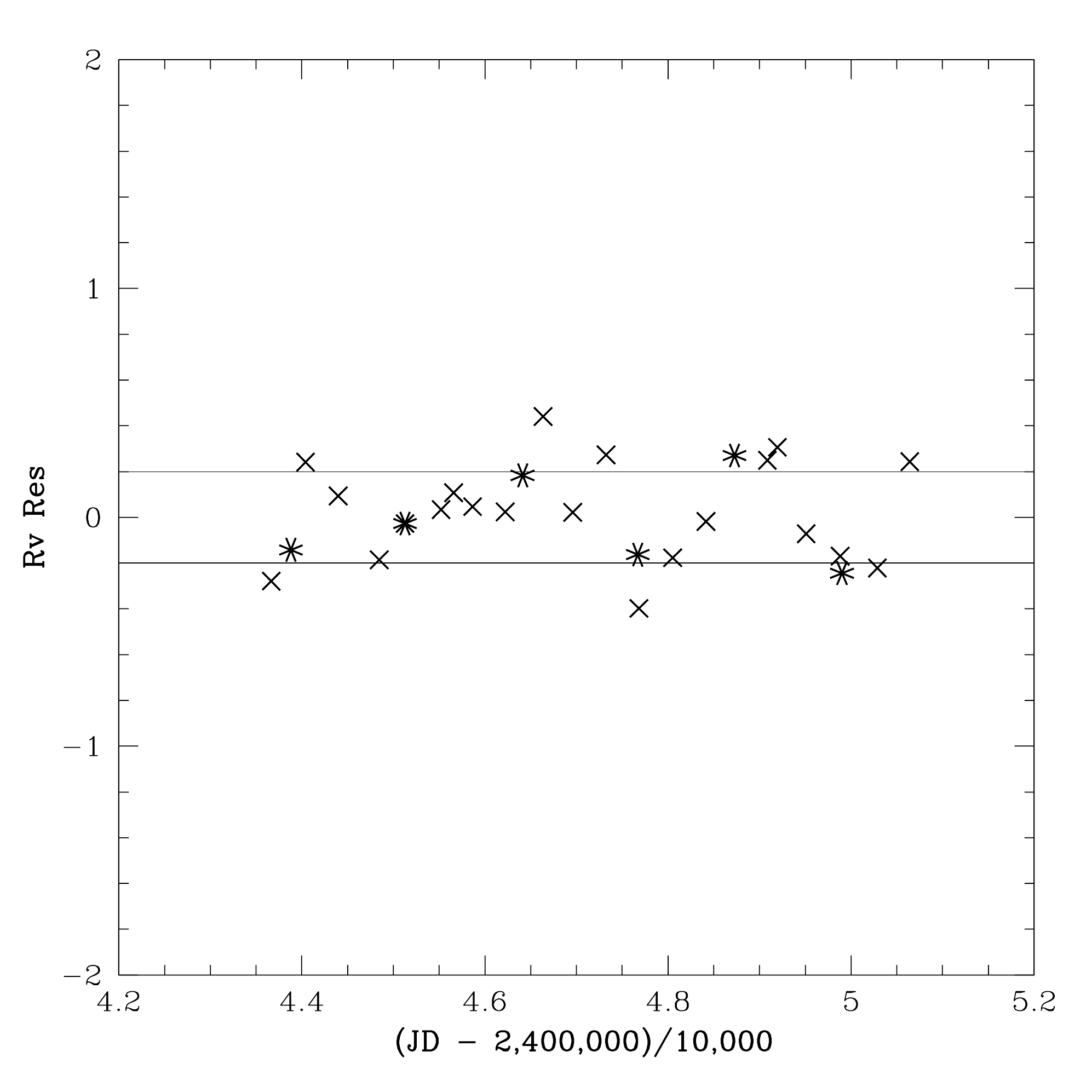}{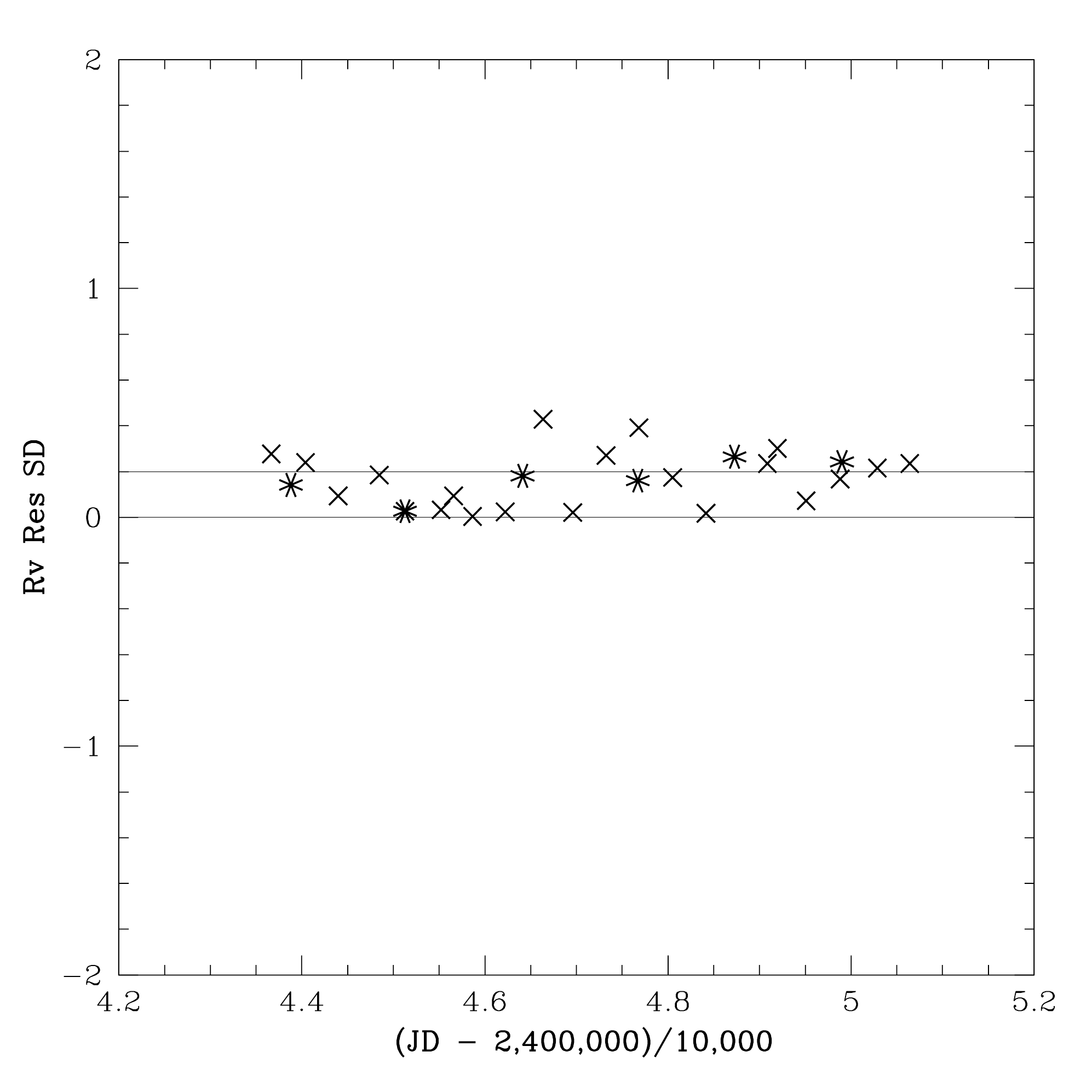}
\caption{Left: Radial velocities binned into a year (x) and in 
periods of low amplitude (*).  Each bin is plotted using the 
first JD of the bin. Lines indicate $\pm$ 0.2 km s$^{-1}$.  Right: 
Standard deviation of each bin.  Symbols are the same as left plot.  
\label{vr.burki.mean}}
\end{figure}

A companion is  surprising as there 
was no previous evidence that the Cepheid is a member of 
a spectroscopic binary. 
Identifying binary properties in a V473 Lyr system 
is, of course, complicated by variations due to pulsation, 
including the variable amplitude. 
MS discuss possible orbital velocity variation based on the 
velocities of Burki (2006) in connection with a third (5290$^d$)
modulation in V473 Lyr.  
Fig.~\ref{vr.burki} shows the radial 
velocities over nearly 40 years. Velocities before JD 2452000
are CORAVEL velocities  as discussed by Burki (2006, kindly made
available
by the Geneva Observatory), listed in  
Appendix C.  
More recent velocities are from Molnar, et al. (2017) and 
Appendix B.  The bottom 
panel in Fig.~\ref{vr.burki} shows the residuals after 
correction for pulsation including modulation, based on  
the analysis of MS.  

Residuals in the lower panel in  Fig~\ref{vr.burki} sometimes
 exceed the typical CORAVEL error of 1 km s$^{-1}$
(e.g. Evans, et al. 2015), although the scatter in measures obtained
on the same night (Appendix C) suggest 1 km s$^{-1}$ is a reasonable estimate.  
In order to look for multi-year velocity variations which might 
come from orbital motion, we have explored the data in two ways. 
First, we have binned data into one year segments to look at 
the mean and standard deviation. (Since the velocity uncertainties
in the Appendix are all very similar, we have used no weighting.) 
  Fig.~\ref{vr.burki.mean}
 shows the mean and standard deviation 
in these bins.  The means are typically  within 
$\pm$0.2 km s$^{-1}$, which is consistent with the standard 
deviations within each bin.

The second approach is as follows.  Since the amplitude of the pulsation  is 
very small for some periods, corrections for pulsation should also
be very small.  We have identified periods of low amplitude and 
created the means and standard deviations in those bins.  As shown 
in   Fig.~\ref{vr.burki.mean}  by 
asterisks, the means and sd at these modulation phases are very 
similar to the results for the annual means.  This demonstrates that the 
corrections for pulsation are accurate (for both high and low 
amplitude phases) and confirms the values for the annual means.  
In summary  Fig.~\ref{vr.burki.mean} 
provides an estimate of velocity variations of 0.2 km s$^{-1}$ 
due to orbital 
motion over the 17 year coverage.



\begin{figure}
\plotone{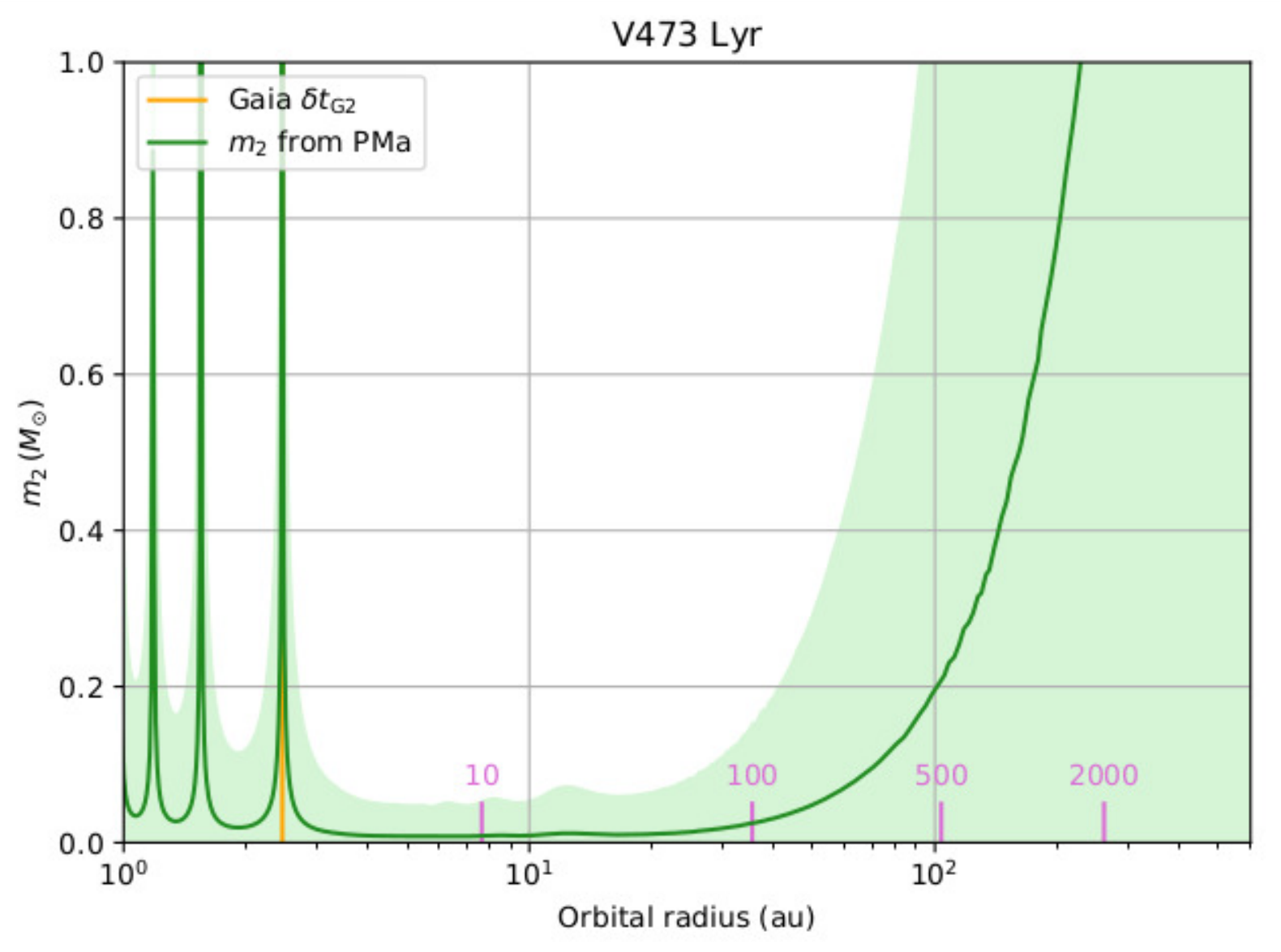}
\caption{The constraints on the orbit from {\it Gaia} 
proper motions.  X-axis: orbital radius in au and orbital
periods (pink) in years; Y-axis: companion mass  
in M$_\odot$; the green curve 
is the relation from the PMa, with the light green shading 
indicating a one $\sigma$ uncertainty; the yellow line is the orbital
radius corresponding to {\it Gaia} DR2 time window of 
688$^d$.  (Plot created using the techniques described in 
Kervella, et al. 2019).
\label{orb.gaia}}
\end{figure}

Comparison of proper motions from {\it Hipparcos} and 
{\it Gaia} (DR2) can reveal orbital motion (proper motion 
anomalies or PMa), as fully discussed by Kervella, et al.
(2019).  Fig.~\ref{orb.gaia} shows the constraints on 
the combination of companion mass M$_2$ and radius 
(using a circular orbit)
found   from the {\it lack} of a significant PMa.
The figure shows that an orbit of 
3 au or one of more than 30 au  would be compatible with 
a companion mass of at least a several tenths of a solar mass, 
but between them a stellar mass companion is unlikely.   
Furthermore, below 3 au there are only the very restricted peaks 
related to aliases of the time window of the {\it Gaia}
DR2 data.  It is unlikely that an orbital period would 
happen to fall in these limited regions.

\begin{figure}
\plotone{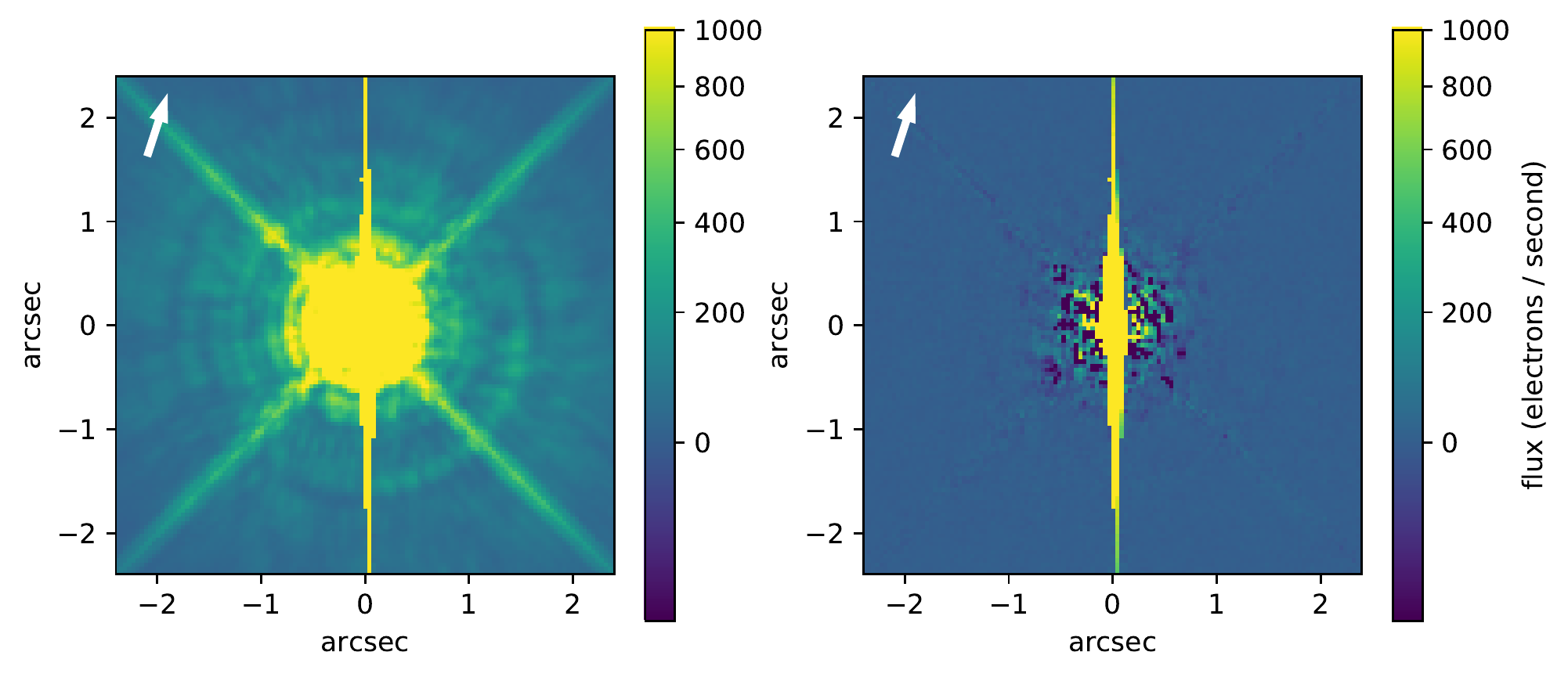}
\caption{The HST WFC3 image of V473 Lyr.  Left: Original image.
Right: Point spread corrected image.  The arrow points North.
\label{clos.comp}}
\end{figure}

  Based on the proper motion constraint 
on orbits, we can estimate possible orbital velocities
for the V473 Lyr system.  For the 
Cepheid we use a mass of 4 M$_\odot$, based on the mass of V1334 Cyg
(Gallenne, et al. 2018).  The X-ray luminosity of the companion 
is consistent with a G star at the age of the Cepheid 
(Preibisch and Feigelson 2005).  
 There is some scatter in the age--spectral type--luminosity
relations, but 1 M$_\odot$ is a reasonable estimate for the companion.
A semi-major axis of 3 au corresponds to an orbital period
of 2.3 years with these values. Note that 
orbits with Cepheid primaries all have periods of a year or more 
(Evans, et al. 2015) since shorter period systems would have 
undergone Roche lobe overflow when the primary became a red
giant.  For Cepheid masses, an orbit with a period of a year corresponds
to an orbital radius of 1.7 au.
  A simple  circular 
orbit with a period of 2.3 years 
would have a Cepheid velocity of 10 km s$^{-1}$ 
and companion velocity of 39 km s$^{-1}$. These numbers will 
be reduced, of course, if the orbit is inclined or 
eccentric.  Comparable velocities for a 30 au orbit (0.05$"$) with
a 73 year period become 2.7 km s$^{-1}$ for the 
Cepheid and 11  km s$^{-1}$ for the companion.  The velocity 
data (Fig.~\ref{vr.burki} and Fig.~\ref{vr.burki.mean}) 
make the short period orbit very unlikely, 
requiring a very small inclination. This is 
particularly true since  a period of nearly 20 years is 
very well covered by the Burki data.  An orbit of 30 years or longer remains
possible.

There is also a  constraint on the 
outer size of the orbit.  Seventy Cepheids 
have been surveyed with the HST Wide Field Camera 3 (WFC3)
to detect resolved companions (Evans, et al. 2016b).  The final
analysis uses point spread corrected images as described 
in Evans, et al. 2020).  On the corrected images, 
companions as close as 0.5$\arcsec$ can be identified (Fig.~\ref{clos.comp}).  
No such companions are detected to this limit for
V473 Lyr.  Using the distance 
553 pc (Evans et al. 2016b, based on the Benedict, et al. 2007
Leavitt law calibration), this limits the companion to 
within 280 au.  The parallax from {\it Gaia} provides a closer
distance of 453 pc (hence a tighter limit).  
Furthermore, the {\it Gaia} data identified no wide common proper
motion companions.


This identification of a low mass companion has two implications. 
First, low mass companions of Cepheids are difficult to identify, 
since the spectral energy distribution is similar to that of the Cepheid, 
and the companion is much fainter than the Cepheid.  Thus the 
X-ray identification of such a companion is important in 
determining the distribution of mass ratios down to small values.  
Second, as discussed above, we have usually used X-ray observations
to confirm that a resolved companion is  young enough to be
a Cepheid companion.  In this case, the argument is reversed.
The X-ray active companion 
is  another confirmation that V473 Lyr is a young Pop I 
star itself, despite its unusual pulsation characteristics.

\section{Summary}  We have observed V473 Lyr twice with {\it XMM}
with the following results.

$\bullet$ The X-ray flux is reasonably constant through a large part
of the pulsation cycle, implying that a low mass companion 
is the likely source.  

$\bullet$ Limits on the orbit of the companion from 
{\it HST} images, {\it Gaia} proper motions, and radial velocities
are consistent with a separation between 30 and 300 au. 

$\bullet$
This is important both  because 
such a companion is otherwise difficult to detect.

$\bullet$ Furthermore
it confirms that V473 Lyr is a Pop I Cepheid, though an 
unusual one.  

$\bullet$ The X-ray upper limits in Fig~\ref{cep.sum} indicate nine 
other Cepheids where a companion as bright as that for V473 Lye would 
have been detected.

Acknowledgements:  Radial velocities from the Geneva observatory 
were provided by  Stephane Udry and Maxime Marmier.
Support for this work was provided by HST Grant GO-12215.01-A 
and from the {\it Chandra} X-ray Center NASA Contract NAS8-03060.
Support was also provided by the Lend\"ulet Program of the 
Hungarian Academy of Sciences, project No. 2018-7/2019.
L.M. was supported by the Premium Postdoctoral Research Program 
of the Hungarian Academy of Sciences.
E.~P. was supported by the J\'anos Bolyai Research Scholarship 
of the Hungarian Academy of Sciences and by  the Hungarian 
National Research, Development and Innovation Office (NKFIH) grant PD-121203.
Support for HMG was provided by the National Aeronautics and Space 
Administration through the Smithsonian Astrophysical Observatory 
contract SV3-73016 to MIT for Support of the Chandra X-Ray Center, 
which is operated by the Smithsonian Astrophysical Observatory for 
and on behalf of the National Aeronautics Space Administration under 
contract NAS8-03060.
P.K. acknowledges the support of the French Agence Nationale de la 
Recherche (ANR), under grant ANR-15-CE31-0012-01 (project UnlockCepheids).
The research leading to these results  has received funding from the European 
Horizon 2020 research and innovation programme (grant agreement No 695099).
This discussion is based on observations obtained with {\it XMM-Newton},
an ESA science mission with instruments and contributions 
directly funded by ESA member states and NASA.  
It is based in part in observations made with the NASA/ESA
{\it Hubble Space Telescope} obtained by the Space Telescope Science
Institute. STScI is operated by the Association of Universities
for Research in Astronomy Inc., under NASA contract NAS5-26555.
This  work  has  made  use  of  data  from  the  European
Space  Agency  (ESA)  mission Gaia (http://www.cosmos.esa.int/gaia),
processed  by  the Gaia Data  Processing  and  Analysis  Consortium  
(DPAC, http://www.cosmos.esa.int/web/gaia/dpac/consortium). 
Funding for the DPAC has been provided by national institutions, 
in particular the institutions  participating  in  the Gaia
Multilateral  Agreement.  SIMBAD was used in the preparation 
of this paper.




\section{Appendix A: Photometry}

Table 2 lists the photometry from 2019
 Only the first few lines are provided here to show the 
scope of the data; the full dataset is included in the 
electronic version.

\begin{deluxetable}{lrc}
\tablecaption{Photometry of V473 Lyr  \qquad\qquad\qquad\qquad \label{}}
\tablewidth{0pt}
\tablehead{
\colhead{JD} &  \colhead{V}  & \colhead{sd}  \\
 \colhead{-2,400,000}  &  \colhead{mag} &  \colhead{mag}  \\
}
\startdata
58553.5642  & 	6.209 & 	0.001 \\
58553.5644 & 	6.211 & 	0.001 \\
58553.5646 & 	6.166 & 	0.001 \\
58553.5647 & 	6.208 & 	0.001 \\
58553.5649 & 	6.192 & 	0.001 \\
58553.5651 & 	6.159 & 	0.001 \\
58553.5652 & 	6.192 & 	0.001 \\
58553.5654 & 	6.170 & 	0.001 \\
58553.5656 & 	6.172 & 	0.001 \\
\enddata

\end{deluxetable}

\section{Appendix B: Velocities (2019)}

Table 3 lists the radial velocities from 2019
 Only the first few lines are provided here to show the 
scope of the data; the full dataset is included in the 
electronic version.

\begin{deluxetable}{lrc}
\tablecaption{Radial Velocities of V473 Lyr in 2019 \qquad\qquad\qquad\qquad \label{}}
\tablewidth{0pt}
\tablehead{
\colhead{JD} &  \colhead{V$_R$}  & \colhead{sd}  \\
 \colhead{-2,400,000}  &  \colhead{km s$^{-1}$} &  \colhead{km s$^{-1}$}  \\
}
\startdata
58559.580   & 	-14.386 & 	0.605 \\
58559.587 &  	-14.157 & 	0.846 \\
58559.595 & 	-13.974 & 	0.553 \\
58559.602 & 	-14.046 & 	0.553 \\
58559.610 & 	-13.850 & 	0.641 \\
58559.617 & 	-13.939 & 	0.562 \\
\enddata

\end{deluxetable}

\section{Appendix C: CORAVEL Velocities}

Table 4 lists the CORAVEL velocities discussed by Burki (2006).
 Only the first few lines are provided here to show the 
scope of the data; the full dataset is included in the 
electronic version.

\begin{deluxetable}{lrc}
\tablecaption{CORAVEL Velocities of V473 Lyr  \qquad\qquad\qquad\qquad \label{}}
\tablewidth{0pt}
\tablehead{
\colhead{JD} &  \colhead{V$_R$}  & \colhead{sd}  \\
 \colhead{-2,400,000}  &  \colhead{km s$^{-1}$} &  \colhead{km s$^{-1}$}  \\
}
\startdata
43286.527 &	 -14.81	 & 0.39 \\
43666.609 &	 -17.99	 & 0.49 \\
43706.471 &	 -25.24	 & 0.37 \\
43708.471 &	 -13.04	 & 0.38 \\
43709.449 &	 -25.05	 & 0.39 \\
43709.522 &	 -24.93	 & 0.44 \\
43710.472 &	  -7.89	  & 0.38 \\
43711.474 &	 -12.56	 & 0.37 \\
43712.452 &	 -24.87	 & 0.38 \\
\enddata

\end{deluxetable}

\end{document}